

\input harvmac

\Title{BROWN-HET-915, hep-th/9307122}{Dirichlet Strings}
\centerline{Miao Li}
\bigskip
\centerline{Department of Physics}
\centerline{Brown University}
\centerline{Providence, RI 02912}
\centerline{li@het.brown.edu}
\bigskip
Strings propagating along surfaces with Dirichlet boundaries are
studied in this paper. Such strings were originally proposed as a possible
candidate for the QCD string. Our approach is different from previous
ones and is simple and general enough, with which basic problems can be
easily addressed. The Green function on a surface with Dirichlet boundaries
is obtained
through the Neumann Green function on the same surface, by employing a simple
approach to Dirichlet conditions. An easy consequence of the
simple calculation of the Green function is that in the simplest model,
namely the bosonic Dirichlet string, the critical dimension is still 26,
and the tachyon is still present in the spectrum, while the scattering
amplitudes differ dramatically from those in the usual string theory. We
discuss the high energy, fixed angle behavior of the four point
scattering amplitudes on the disk and the annulus. We argue for general
power-like behavior of arbitrary high energy, fixed angle scattering
amplitudes. We also discuss the high
temperature property of the finite temperature partition function on an
arbitrary surface, and give an explicit formula of the one on the annulus.

\Date{7/93}

\newsec{Introduction}

A theory of the QCD string is still not in our sight, despite many efforts
in years. The topological expansion in terms of $1/N$ \ref\thooft
{G. 't Hooft, Nucl. Phys. B72 (1974) 461; B75 (1974) 461.}  \ref\vene{G.
Veneziano, Nucl. Phys. B74 (1974) 365.} and Regge phenomenology strongly
suggest that such a theory exists. More recently, 2D pure Yang-Mills theory has
been successfully interpreted as a string theory in terms $1/N$ expansion
\ref\qcd{D.J. Gross, preprint PUPT-1356 (1992); D.J. Gross and W. Taylor,
preprint PUPT-1376/LBL-33458 (1993), PUPT-1382/LBL-33767 (1993);
J. Minahan, preprint UVA-HEP 92010 (1993).}.
Many proposals and attempts in constructing the 4D QCD string were reviewed
in \ref\polch{J. Polchinski, Strings and QCD?, talk at the Symposium on
Black Holes, Wormholes, Membranes, and Superstrings, H.A.R.C., Houston,
1992.}.

Before trying on any attempt, one should have in mind how different
the fundamental string theory is from a putative QCD string. First of
all, many massless states in the spectrum of the usual string theory
are not present in QCD, as there must be a mass gap in QCD. One might hope
that radiative corrections may elevate these massless states to be massive.
Another salient feature of the fundamental string theory not shared
by QCD is the extreme softness in high energy scattering amplitudes.
Recently, Polchinski calculated the string free energy per unit length
\ref\polchs{J. Polchinski, Phys. Rev. Lett. 68 (1992) 1267.}.
The result is dramatically different from that in string theory. A
consequence is that to achieve that form of free energy, one would require
the number of degrees of freedom on the world sheet increase
with temperature indefinitely. Both the high energy and high temperature
behavior of QCD hint at point-like structure. It is desirable to have a string
theory interpolating from low energy Regge behavior to high energy point-like
structure.

M. Green has proposed a simple modification of the usual string theory,
in the way that one adds Dirichlet boundaries to the world sheet. Since
these boundaries are mapped to single spacetime points, intuitively
point-like structure may appear at high energies. He actually calculated
some four-point amplitudes on the disk with a Dirichlet boundary, and obtained
power fall-off at high energy \ref\mbg{M.B. Green, Phys. Lett. B266
(1991) 325.}. He also calculated mass shift of a winding tachyon state
on the disk at finite temperature
\ref\mbgg{M.B. Green, Phys. Lett. B282 (1992) 380.}. He obtained
a similar result as predicted by Polchinski for QCD. These results show
that the Dirichlet string indeed has point-like structure at high energies.
We find this model fascinating, because it is simple enough to be useful
in doing actual calculations. Green's approaches in above cited papers
are rather inconvenient, and it is hard to generalize them to do general
calculations.

There is an alternative model to the Dirichlet string theory, suggested
in \ref\zg{Z. Qiu, Phys. Lett. B306 (1993) 261.}. In that theory, the world
sheet has
no boundary. Instead mapping the boundary to spacetime points, part of the
world sheet is mapped to spacetime points. We will not discuss this
interesting model here.

It is the aim of this paper to present a simple and general approach to the
Dirichlet string. With this approach, we are able to reproduce Green's
results, and obtain some new results on the annulus. These results can be
easily generalized to other cases. With our approach,
it is straightforward to employ modern conformal field theory techniques
to do calculations. Some basic problems such as determination of the critical
dimension can be easily addressed.

We shall present our simple approach to the Dirichlet string theory in
sect. 2. The Green function is related to the Neumann Green function
on the same surface. Two examples are given. These Green functions on the
disk and the annulus will be used later. We will also argue that the Weyl
anomaly takes the same form as in the Neumann theory, therefore the critical
dimension is 26, order by order in the perturbation theory. We show that
the divergence arising from integration over moduli on the annulus is the
same as in the Neumann theory, therefore there is no Fischler-Susskind
mechanism available to modify the critical dimension.

Scattering amplitudes are discussed in sect. 3. We shall reproduce Green's
result on the disk. We also consider the annulus. In case of high energy,
fixed angle scattering, the major contribution to the amplitude arises when
all vertex operators are on the boundary. Power-like behavior then emerges
generally. The underlying mathematical reason is the following. When vertex
operators are all on the boundary of any surface, may be on different
components, the Green
function becomes independent of positions of these vertex operators. Off the
boundary, integrand in the integral expression of the amplitude oscillates
fast. This situation is very different from that in the usual string theory,
where saddle points usually sit in the middle of the moduli space. The
physical picture of high energy, fixed angle scattering is as follows. When
all high energy particles approach together, they suddenly are forced to
interact at points where the Dirichlet boundaries are mapped to. When there
are, say, two such points, then the world sheet connecting these point
looks more like a usual Feynman diagram. This property is one of
serious reasons that the Dirichlet string is a candidate for the QCD string.
The physical picture we described above is very similar to that in QCD
\ref\qcds{see articles in Perturbative Quantum Chromodynamics, ed. A.H.
Mueller, World Scientific, 1989.}. In QCD, when two high energy hadrons
scatter at fixed angle,
they exchange many soft gluons before they come near each other. They
suddenly exchange hard gluons when they approach together, this exhibits
point-like interaction. After they depart from each other, they exchange
soft gluons again.

We discuss finite temperature string free energy in sect. 4. We study
general properties of the free energy at high temperature. It is rather
striking that nonvanishing winding modes along components of the boundary
contribute. So the sphere free energy has temperature dependence. This
looks contradictory to the fact that no matter how many windows we open on
the sphere, they are mapped to spacetime points. There is no such
contradiction. As we shall show by explicit calculating the annulus free
energy, there is divergence on the boundary suppressing nonvanishing winding
modes. This divergence can be removed by adding counter term on the boundary.
When this is done, the temperature dependent part becomes finite.
We consider such a regularization provides us a mechanism of opening
little windows on the sphere, which is what one expects from studying large
$N$ QCD \ref\thorn{C. Thorn, Phys. Lett. B99 (1981) 458.}. Its significance
deserves further study.

We present discussion on our results in sect. 5.

\newsec{General properties}

The basic quantity needed in calculating scattering amplitudes and other
physical observables is the Green function on a surface with Dirichlet
boundaries. We shall introduce in the next subsection a simple approach
to the model, namely introducing a boundary field to effectively impose on
Dirichlet conditions. The modified Green function is then easily obtained,
provided the Green function with Neumann boundary conditions is known.
We demonstrate this calculation on the disk and the annulus. With help of
this modified Green function, we shall show in a subsequent subsection that
the Weyl anomaly is essentially the same as for the surface with Neumann
boundaries, so that the critical dimension of the Dirichlet string is
also 26. We also argue that the moduli dependence of the partition function
on the annulus is the same as the one for the Neumann theory, thus there is no
analogue of Fischler-Susskind mechanism to modify the critical dimension
``quantum mechanically''.

\subsec{The Green function}

The Riemann surface under consideration has  $L$ components of boundary, we
use $\theta_i$ to parameterize the i-the component. A metric $g_{ab}$
on the Riemann surface is assumed, with induced line element $ds_i$ on each
boundary. We consider only bosonic string in this paper. There are
$D$ scalars $X^{\mu}$ as coordinates of D-dimensional spacetime living on the
surface. In the Polyakov path integral prescription, we first integrate over
all possible configurations $X^\mu (z)$ with Dirichlet conditions
$X^{\mu}(\theta_i)
=x^{\mu}_i$, then integrate out $x_i$ on each component. The simplest way
to calculate the Green function is to consider the correlation
function of vertex operators $\hbox{exp}(ip_iX(z_i))$. With the path integral
prescription, the correlation is defined by
\eqn\start{\langle \prod_{i=1}^n e^{ip_iX(z_i)}\rangle=
\int \prod_{i=1}^Ld^Dx_i[dX]e^{-S(X)}\prod_{i=1}^n e^{ip_iX(z_i)},}
where the world sheet action is
$$S(X)={1\over 8\pi}\int \sqrt{g}g^{ab}\partial_aX\partial_bX d^2z.$$
Spacetime index is always suppressed in this paper.
Insertions of vertex operators in \start\ can be absorbed into the action,
provided we rewrite
$$\prod_{i=1}^{n}e^{ip_iX(z_i)}=\hbox{exp}\left(i\int
\sqrt{g}d^2zX(z)\sum_ip_i\delta^2
(z-z_i)\right).$$
The path integral in \start\ is Gaussian, and can be performed.
A usual way to do this is to decompose $X(z)$ into the classical solution
and the fluctuation piece, $X(z)=X_{cl}(z)+\xi(z)$. $\xi(z)$ will decouple from
$X_{cl}(z)$ and can be integrated out. With fixed $x_i$, the classical solution
satisfies
$$\eqalign{-\Delta X_{cl}(z)&=4\pi i\sum_ip_i\delta^2(z-z_i),\cr
X_{cl}(\theta_i)=x_i,}$$
where $\Delta$ is the Laplacian associated with metric $g$. If $X_{cl}$ can
be readily found, then our problem is solved. However, there
are two problems with this approach. First, we are not sure if classical
solution of this type for arbitrary $x_i$ always exists (if it does, it is
unique). Second, even it always exists, integral over $x_i$ is hard to
perform. To bypass these problems, we shall adopt another approach.
To effectively impose on Dirichlet boundary conditions, we insert delta
functions
$\delta(X(\theta_i)-x_i)$ into \start. These delta functions in turn are
written
by introducing fields $\rho_{\mu}(\theta_i)$ living only on the boundary:
$$\delta(X(\theta_i)-x_i)=\int[d\rho(\theta_i)]\hbox{exp}\left(-{i\over 4\pi}
\int \rho(\theta_i)(X(\theta_i)-x_i)ds_i\right).$$
Upon insertion of this representation of the delta function, we perform
integral over $x_i$ directly. This results in constraint on $\rho$,
$\int \rho(\theta_i)ds_i=0$, that the $\rho$ fields contain no constant
modes. Finally, we have the following path integral definition of the
correlation function
\eqn\def{\langle \prod_{i=1}^ne^{ip_iX(z_i)}\rangle=\int [dX][d\rho]'
e^{-S(X,\rho)}\prod_{i=1}^ne^{ip_iX(z_i)},}
where $[d\rho]'$ denote path integral of fields $\rho(\sigma_i)$ with constant
modes omitted. Note that the boundary conditions for $X(z)$ are free, $X$
takes arbitrary values on components of the boundary.
The action in \def\ reads
\eqn\act{S(X,\rho)={1\over 4\pi}({1\over 2}\int\sqrt{g}g^{ab}\partial_aX
\partial_bX
d^2z+i\sum_{i=1}^L\int\rho(\theta_i)X(\theta_i)ds_i).}
To perform the Gaussian path
integral over $X$ in \def, we separate $X(z)$ into the classical piece and the
fluctuation piece, $X(z)=X_{cl}(z)+\xi(z)$. The classical solution satisfies
\eqn\eom{\eqalign{-\Delta X_{cl}(z)&=4\pi i\sum_i p_i\delta^2(z-z_i),\cr
\partial_nX_{cl}(\theta_i)&=-i\rho(\theta_i),}}
where $\partial_n$ is the normal derivative on the boundary.
This classical solution is obtained through Green functions $G_N(z,w)$
and $\chi(z,\theta_i)$. They satisfy
\eqn\neu{-\Delta G_N(z,w)=\delta^2(z,w),\quad
\partial_nG_N(\theta_i, w)=c_i,}
and
\eqn\boun{-\Delta\chi(z, \theta_i)=0,\quad \partial_n\chi(\theta_i,
\theta_j)=\delta_{ij}\delta(\theta_i,\theta_j).}
Two remarks concerning these equations are in order. First, $c_i$ are
constants in \neu. They cannot be all set to zero, since by Gauss theorem,
$\sum_i\int \partial_nG_N(\theta_i, w)ds_i=-1$. As long as this condition
is fulfilled, the Neumann function $G_N(z,w)$ always exists. Different
choices of $c_i$ give the same answer in our calculation, since the energy
momentum is conserved
$\sum_i p_i=0$. Second, the Green function $\chi(z, \theta_i)$ is associated
with the boundary source $\rho(\theta_i)$. This field has no constant mode,
so we would like to impose $\int \chi(z,\theta_i)ds_i=0$. Also, the delta
function in \boun\ contains no constant mode. With help of these Green
functions, the classical solution is obtained
\eqn\class{X_{cl}(z)=4\pi i\sum_{i=1}^np_iG_N(z,z_i)-i\sum_{i=1}^L
\int\chi(z, \theta_i)\rho(\theta_i)ds_i.}
Substituting $X=X_{cl}+\xi$ into the path integral, we find the action
including sources is separated into two pieces $S_{cl}+S(\xi)$. $S(\xi)$
is independent of $p_i$ and $\rho$, so path integral over fluctuations $\xi$
yields a factor depending only on geometry, which is irrelevant for
our calculation of the Green function. The classical solution $X_{cl}$ is
unique up to a constant. Upon integration over this zero mode, we get
constraint $\sum_ip_i=0$. We shall assume this condition be always met.
Then the classical part of the action, with $X_{cl}$ in \class, reads
\eqn\cac{\eqalign{S_{cl}&=-{1\over 2}\sum_{i,j}p_j\int\rho(\theta_i)
G_N(\theta_i,
z_j)ds_i+2\pi\sum_{i,j}p_ip_jG_N(z_i,z_j)\cr
&+{1\over 8\pi}\sum_{i,j}\int \rho(\theta_i)\chi(\theta_i,\theta_j)
\rho(\theta_j)ds_ids_j-{1\over 2}\sum_{i,j} p_j\int\chi(z_j,\theta_i)
\rho(\theta_i)ds_i.}}
It can be proved that the last term in the above formula is equal to the first
term.
More than this, we show that $\chi(z,\theta_i)$ is expressible in terms
of $G_N(z,\theta_i)$. Let $l_i=\int ds_i$, the circumference of the i-th
component of the boundary, then
\eqn\use{\chi(z,\theta_i)=G_N(z,\theta_i)-{1\over l_i}
\int G_N(z,\theta_i)ds_i.}
Note that $\chi(z,\theta_i)$ is essentially the same as $G_N(z,\theta_i)$,
except that the constant mode is subtracted. Formula \use\ can be proved
by considering quantity
$$\int\chi(w,\theta_i)(-\Delta_wG_N(w,z))d^2z.$$
It is just $\chi(z,\theta_i)$, by using \neu. On the other hand, we do
integration
by part and obtain $\sum_j\int\partial_n\chi(\theta_j,\theta_i)G_N(z,\theta_j)
ds_j$. Using \boun\ and noting that there is no constant mode in the boundary
delta function, we obtain \use. It follows from \use\ that the first term
in \cac\ is equal to the last term in \cac.

The final step in obtaining the
correlation function in \def\ is to integrate out fields $\rho(\theta_i)$.
Again the dependence of \cac\ on $\rho$ is quadratic. To complete the
square, we need the inverse of $\chi(\theta_i,\theta_j)$:
$$\sum_k \int\chi(\theta_i, \theta_k)\chi^{-1}(\theta_k, \theta_j)ds_k=
\delta_{ij}\delta(\theta_i,\theta_j),$$
again the delta function does not contain the constant mode. Now $S_{cl}$
is written as
\eqn\squ{S_{cl}={1\over 8\pi}\sum_{ij}\int\tilde{\rho}(\theta_i)\chi(
\theta_i,\theta_j)\tilde{\rho}(\theta_j)ds_ids_j+2\pi\sum_{ij}
p_ip_jG(z_i,z_j),}
where
\eqn\green{G(z_i,z_j)=G_N(z_i,z_j)-\sum_{k,l}\int\chi(z_i,\theta_k)\chi^{-1}
(\theta_k,\theta_l)\chi(z_j,\theta_l)ds_kds_l,}
and
$$\tilde{\rho}(\theta_i)=\rho(\theta_i)-4\pi\sum_{j,k}p_k\int\chi^{-1}(
\theta_i,\theta_j)\chi(z_k,\theta_j)ds_j.$$
Just as for $\chi(\theta_i,\theta_j)$, we require its inverse $\chi^{-1}$
contain no constant mode. From the definition of $\tilde{\rho}$, we see that
this function contains no constant mode too, therefore we can replace path
integral measure $[d\rho]'$ in \def\ by $[d\tilde{\rho}]'$. Integration
over $\tilde{\rho}$, according to \squ, gives us a factor independent of
$p_i$. Finally, we obtain the correlation function
\eqn\resul{\langle\prod_ie^{ip_iX(z_i)}\rangle=Ne^{-2\pi\sum_{ij}p_ip_j
G(z_i,z_j)},}
where the Green function $G(z_i,z_j)$ is given in \green. $N$ is a factor
depending on geometry only. We will discuss this factor in subsect. 2.3
where we discuss Weyl anomaly. Note that, when $i=j$ in \resul, divergence
occurs, this divergence is absorbed into a rescaling of the vertex operator.
The propagator $\langle X(z)X(w)\rangle$ is readily read off from \resul:
\eqn\propa{\langle X^{\mu}(z)X^{\nu}(w)\rangle=4\pi\eta^{\mu\nu} G(z,w).}
This is our main result in this subsection. What could have been boundary
conditions for $G(z,w)$? From \green\ and \use, we find $G(z,\theta_i)=1/l_i
\int G_N(z,\theta_i)ds_i$. So when one argument is on the boundary, $G$ is
independent of the position of this argument on the boundary.

\subsec{Two examples}

We have shown that the Green function with Dirichlet conditions can be obtained
from the Neumann Green function. We demonstrate this calculation by two
examples, the disk and the annulus. The Green functions of these two geometries
will be used in our subsequent calculations.

\noindent{\it A. The disk}

Consider the unit disk. There is no moduli parameter for this geometry.
The Neumann Green function is
\eqn\dis{G_N(z,w)=-{1\over 4\pi}(\hbox{ln}|z-w|^2+\hbox{ln}|1-z\bar{w}|^2),}
where the first term is the usual Green function on the sphere, and the second
term is due to a mirror charge. When $w$ is on the boundary $w=\hbox{exp}(i
\theta)$, $G_N(z,\theta)$ contains no constant mode, so $\chi(z,\theta)
=G_N(z,\theta)$. As a check, $\partial_r\chi(\theta,\theta')=
1/(2\pi)\sum_{n\ne 0}\hbox{exp}(in(\theta-\theta'))$, just the delta function
without the constant mode. The inverse of $\chi(\theta,\theta')$ is given by
$\chi^{-1}(\theta,\theta')=\partial_\theta\partial_{\theta'}
\chi(\theta,\theta')$. Concretely
$$\chi^{-1}(\theta,\theta')={1\over 2\pi}\sum_{n\ge 1}ne^{in(\theta-\theta')}
+c.c.$$
With this inverse, $G(z,w)$ can be easily calculated, using \green. The
result is
\eqn\disk{G(z,w)=-{1\over 4\pi}(\hbox{ln}|z-w|^2-\hbox{ln}|1-z\bar{w}|^2).}
It is amazing to notice that the image part in $G(z,w)$ has a different sign
from that in $G_N(z,w)$. When one of the arguments is on the boundary,
$G=0$. This has important consequences for scattering amplitudes.

\noindent{\it B. The annulus}

There is a real moduli of geometries on the annulus. If we put this annulus
on the complex plan and have the radius of the outer circle to be unit, then
the moduli parameter is the radius of the inner circle, denote it by $a$.
The Neumann Green function is \ref\hsv{C.S. Hsue, B. Sakita and M.A. Virasoro,
Phys. Rev. D2 (1970) 2857.}
\eqn\ann{\eqalign{G_N(z,w)&=-{1\over 4\pi}[\hbox{ln}|z-w|^2+
\hbox{ln}|1-z\bar{w}|^2
+\sum_{n=1}^\infty(\hbox{ln}|1-a^{2n}{z\over w}|^2|1-a^{2n}{w\over z}|^2\cr
&+\hbox{ln}|1-a^{2n}z\bar{w}|^2|1-a^{2n}{1\over z\bar{w}}|^2)].}}
It is easy to see that there is a constant mode in $G_N$ in this case, when
one argument is on the inner circle, so $\chi$ is different from $G_N$ by
this term. We denote by $\chi_{11}(\theta,\theta')$ the Green function $\chi$
when both arguments are on the outer circle, by $\chi_{22}(\theta,\theta')$
when both arguments are on the inner circle, and by $\chi_{12}(\theta,
\theta')$ when one argument is on the outer circle and the other is on the
inner one. It is interesting to check that
$$\eqalign{\chi^{-1}_{11}(\theta,\theta')&=\partial_\theta\partial_{\theta'}
\chi_{11}(\theta,\theta'),\quad \chi^{-1}_{22}(\theta,\theta')=\partial_\theta
\partial_{\theta'}\chi_{22}(\theta,\theta'),\cr
\chi^{-1}_{12}(\theta,\theta')&=-\partial_\theta\partial_{\theta'}\chi_{12}
(\theta,\theta')},$$
quite similar to the case of the disk. Note that the minus sign in the third
formula is due to the opposite orientations of the two boundaries. We do not
bother to write down the explicit expressions of these functions. Substitute
these functions into \green, we find
\eqn\annu{\eqalign{G(z,w)&=-{1\over 4\pi}[\hbox{ln}|z-w|^2-\hbox{ln}|1-z
\bar{w}|^2+\sum_{n=1}^\infty (\hbox{ln}|1-a^{2n}{z\over w}|^2|1-a^{2n}{w
\over z}|^2\cr
&-\hbox{ln}|1-a^{2n}z\bar{w}|^2|1-a^{2n}{1\over z\bar{w}}|^2)].}}
Once again we notice that $G$ is almost the same as $G_N$ except for a couple
of sign changes.

We have demonstrated in this subsection that once the Neumann Green function is
given, it is straightforward to calculate the Green function we need.

\subsec{The Weyl anomaly and the critical dimension}

The critical dimension of the Dirichlet string theory has not been determined
previously in an obvious way. It was conjectured in \mbgg\ that the critical
dimension is modified, albeit by some nonperturbative effects. Making use of
our result on the Green functions, we shall show that the critical dimension
is still 26, perturbatively.

First of all, we note that the boundary term in \act\ involving $\rho$
can be made invariant
under metric rescaling, this is achieved by letting $\rho$ transform against
$ds$ under the Weyl rescaling. Because the
ultralocal nature of the boundary term, the only ultralocal anomaly, invariant
under diffeomorphism, should be the boundary cosmological term $\mu\int ds$.
This term is easily removed by adding such a counter term, just as one
usually does with the bulk cosmological term.

There is a easier way to demonstrate the Weyl anomaly than calculating the
path integral directly. The method is based on the definition of bulk stress
tensor
$$\delta S={1\over 4\pi}\int\sqrt{g}T_{ab}\delta g^{ab}d^2z.$$
The variation of the partition function $Z$ with respect to change of the
metric is
$$\delta\hbox{ln}Z =-{1\over 4\pi}\int\sqrt{g}\langle T_{ab}\rangle \delta
g^{ab}d^2z.$$
If we know how to compute the expectation of $T_{ab}$, we know the metric
dependence of the partition function. In case of the Dirichlet string,
there is an additional boundary term:
\eqn\varia{\delta\hbox{ln}Z=-{1\over 4\pi}[\int\sqrt{g}\langle T_{ab}\rangle
\delta g^{ab}d^2z+\int \langle t_\theta\rangle \delta\sqrt{g_{\theta\theta}}
d\theta ],}
the last term is due to the boundary term in the action. The formula for
$T_{ab}$ is the familiar one. The formula for $t_\theta$ is
\eqn\bst{t_\theta=i\rho(\theta)X(\theta).}
We need to know both the expectation value of the bulk stress tensor and that
of $t_\theta$. To calculate $\langle T_{ab}\rangle$, we evoke formula
$$\delta\langle T_{ab}(z)\rangle =-{1\over 4\pi}\int \sqrt{g(w)}\langle
T_{ab}(z)T_{cd}(w)\rangle\delta g^{cd}(w)d^2w.$$
There could be a boundary contribution to the variation of $\langle T_{ab}
\rangle$. It is vanishing, because the correlation of $T_{ab}$ and $t_\theta$
is zero. Having the above formula, one then proceeds in the usual way
\ref\cardy{see, J.L. Cardy, Lectures given at Les Houches summer session,
1988.} to calculate the variation of the expectation value of the stress
tensor. Given the Green function on the disk as in \disk, we find that
correlations among components of the stress tensor are just the same as in
the Neumann case, since the sign change in the last term in \disk\ does
not change the correlator of the stress tensor. This implies that the
expectation value of the stress tensor
is the same as in the Neumann case. We conclude that the Weyl anomaly,
coming from the bulk contribution, is the old one. Same argument applies
to the annulus, and we believe that this is generally true for any geometry.
This standard Weyl anomaly is canceled by the one from ghost contribution,
if $D=26$. Next we consider the variation of $\langle t_\theta\rangle$:
$$\delta\langle t_\theta\rangle=-{1\over 4\pi}\int\langle t_\theta t_\theta
(\theta')\rangle\delta\sqrt{g_{\theta\theta}(\theta')}d\theta'.$$
The only nontrivial Green function for boundary fields is
$$\langle\rho(\theta)X(\theta')\rangle =-4\pi i\delta(\theta,\theta').$$
The gives us a purely contact term for the correlation of $t_\theta$ with
itself. A consequence of this is that $\langle t_\theta\rangle$ is just
constant, upon regularization. We then conclude from \varia\ that the
boundary contribution of $\langle t_\theta\rangle$ to the Weyl anomaly is
a boundary cosmological constant term.

Having shown that the Weyl anomaly is the same as in the usual open string
theory, we might ask the question that if it is possible to have quantum
effects to modify this anomaly. In other words, does divergence occur in
the moduli integration when the moduli parameters are close to the boundary
of the moduli space, in such a way that to remove this divergence one has to
change the critical dimension? This is well-known Fischler-Susskind mechanism
\ref\fsl{W. Fischler and L. Susskind, Phys. Lett. B171 (1986) 383;
C. Lovelace, Nucl. Phys. B273 (1986) 413.}. In the closed bosonic string,
one-loop divergence gives rise to a cosmological constant; and in the
closed-open string case, divergence on the annulus gives rise to the
contribution of open string fields to equation of motion of closed string
fields \ref\clny{C.G. Callan, C. Lovelace, C.R. Nappi and S.A.
Yost, Nucl. Phys. B288 (1987) 525.}. The latter case is an example of how
boundary effects at higher loops have impact on bulk at lower orders.
In the Dirichlet string theory, divergence also occurs on the annulus when
the inner
radius approach to zero. Our question is, is this divergence different from
that on the annulus in open string theory? The answer is no, so we do not
have new divergence which may force us to modify dimensionality. To show this,
we first calculate the partition function $Z(a)$ as a function of $a$. We
simply borrow a formula derived in \ref\acny{A. Abouelsaoud, C.G. Callan,
C.R. Nappi and S.A. Yost, Nucl. Phys. B280[FS18] (1987) 599.}:
\eqn\fs{{\partial \hbox{ln}Z(a)\over \partial a}={4a\over 1-a^2}\int d^2z
\left({1\over\bar{z}^2}\langle T(z)\rangle +{1\over z^2}\langle\overline{T}
(\bar{z})\rangle\right).}
This formula is valid for both Dirichlet boundaries and Neumann boundaries.
To calculate $Z(a)$ using the above formula, we need to calculate $T$ and
$\overline{T}$. These in turn can be obtained from the Green function in
\annu. Only the first terms in the sum in \annu\ contribute. These terms
are no different from those in the Neumann Green function, so we conclude
that $T$ and $\overline{T}$ are the same as in the Neumann case. Consequently
$$Z(a)=\prod_{n=1}^\infty(1-a^{2n})^{-D},$$
Including contribution from ghosts and taking $D=26$, the integral over
moduli is
$$Z=\int_0^1{da\over a}{1\over\eta^{24}(a^2)},$$
where $\eta(q)=q^{1/24}\prod (1-q^n)$. The above integral
has the same divergences as in the Neumann case. The quadratic divergence is
due to the tachyon, and the logarithmic one is due to the massless dilation.

\newsec{Scattering amplitudes}

Scattering amplitudes are often difficult to calculate explicitly in the
string theory, mainly due to the difficulty of performing integral over the
moduli space. However, better understanding of scattering amplitudes often
enables us to reveal important physics, for example knowledge about the
spectrum can be learned from studying poles in tree amplitudes. As we shall
see, explicit formulas for scattering amplitudes in the Dirichlet string
theory are hard to obtain too. We are then satisfied with knowing their
asymptotic behavior at high energies. If we take this model seriously
as a candidate for the QCD string, we would like to see point-like behavior.
The high energy fixed angle scattering amplitudes should be
power-behaved, rather than fall off exponentially as in the usual string theory
\ref\paul{D.J. Gross and P.F. Mende, Phys. Lett. B197 (1987) 129; Nucl.
Phys. B303 (1988) 407; P.F. Mende, H. Ooguri, Nucl. Phys. B339 (1990) 641.}.
Results presented here concerning disk amplitudes were known to Green \mbg,
although
he used a method based on duality between Neumann and Dirichlet
boundaries, which is less straightforward. Our results on the annulus and
more generally any surface are new, and demonstrate the power of our simple
approach. What will turn out surprising is that the annulus four point
amplitude dominates over the disk one, as it falls off as $s^{-2}$ at fixed
angle, $s$ is one of the Mandelstam variables. In the following, we shall
consider the disk amplitudes first.

\subsec{Disk amplitudes}

The scattering amplitude involving only $n$ tachyons is obtained directly
from \resul,
\eqn\ampl{A_n={1\over \cal{N}}\int\langle\prod_id^2z_ie^{ip_iX(z_i)}
\rangle={1\over\cal{N}}\int\prod_id^2z_ie^{-2\pi\sum_{ij}p_ip_jG(z_i,z_j)}.}
The above formula works only for amplitude on the disk. If the topology
of world sheet is more complex, then one need to integrate over the
moduli space
too. The factor $\cal{N}$ in \ampl\ denotes the volume of the group generated
by conformal Killing vectors. For the disk, the global conformal group
is $SL(2,R)$ of three real parameters.  The integral in \ampl\ is
invariant under this group when all $p_i$ are on-shell. It is then necessary
to take the infinite volume of this symmetry group away from this integral.
The effective way to do this is to fix three integration variables. Before
proceeding further, we shall remark that when $i=j$, $G(z_i,z_i)$ is divergent.
The divergent part must be subtracted and only the $z_i$-dependent finite
part is kept. This finite part in the disk case is $1/(4\pi)\hbox{ln}(1-
|z_i|^2)^2$. So we shall keep in mind that only this part is to be used
whenever $i=j$. Now, the conformal transformation on the disk takes the
form of
$$z\rightarrow {\alpha z +\beta\over\bar{\beta}z+\bar{\alpha}},\quad
|\alpha|^2-|\beta|^2=1.$$
It is easy to check that when all $p_i^2=2$, the integral in \ampl\ is
invariant under this transformation. This on-shell condition is the same
as in the case of Neumann boundary, although the Green function $G$ is
changed. We notice that $G(z_i,z_i)$ gives a factor $(1-|z_i|^2)^{-2}$
in the integrand in \ampl, because $p_i^2=2$. So the integral is divergent
at $|z_i|=1$. This divergence is due to the negative mass square of the
tachyon, and will not occur to other massless and massive states. Notice that
in the case of the Neumann boundary, since the sign for the relevant term
in the Green function is different, there is no such divergence for tachyon.
However, this kind of divergence occurs when the state is massive.

By conformal transformations, any point inside the disk can be brought to
the origin, let us fix $z_1=0$. The residual transformations which keep
origin invariant are those with $\beta=0$, and $|\alpha|=1$. With these
transformations, any point on the disk can be brought to be on the real
axis, let us assume $z_2=x_2$. So we have fixed three integrations parameters
in \ampl. The final step in dividing out the volume of $SL(2,R)$ is to
calculate the Jacobian with our ``gauge fixing''. The Jacobian is $x_2$.
The amplitude \ampl\ now becomes
\eqn\fix{A_n=\int x_2dx_2\prod_{i=3}^nd^2z_ie^{-2\pi\sum_{ij}p_ip_j
G(z_i,z_j)},}
with the understanding that in the exponential in the above formula, $z_1=0$
and $z_2=x_2$. Let us specify to the four point amplitude. Use the
Mandelstam variables
$$s=-(p_1+p_2)^2,\quad t=-(p_1+p_3)^2,\quad u=-(p_1+p_4)^2,$$
we find
\eqn\four{A_4(s,t,u)=\int x_2^{-3}(1-x_2^2)^{-2}(1-|z_3|^2)^{-2}(1-|z_4|
^2)^{-2}dx_2d^2z_3d^2z_4e^{-{\cal E}},}
where the ``energy''  $\cal{E}$ reads
\eqn\bizz{{\cal E}=(s+4)\hbox{ln}{|z_3-z_4|\over |1-z_3\bar{z}_4|}+(t+4)
\hbox{ln}|z_3|{|1-x_2^{-1}z_4|\over |1-x_2z_4|}+(u+4)\hbox{ln}|z_4|{
|1-x_2^{-1}z_3|\over |1-x_2z_3|}.}
A remarkable property of ${\cal E}$ is that when all $x_2$, $z_3$ and $z_4$
are on the boundary, it vanishes and becomes momentum-independent. This
in turn is due to the minus sign in the second term in the Green function
\disk. A similar formula was obtained by Green in \ref\mbggg{M.B. Green,
Phys. Lett. B201 (1988) 42.} using a different method.

In the closed string, the high energy behavior of the four point scattering
amplitudes is governed by the saddle point sitting in the middle of the moduli
space \paul. It being so, the energy can not lose its dependence on
momenta. This is the reason why the high energy, fixed angle scattering
amplitude falls off exponentially. Now in our theory, there is a whole
submanifold on the moduli space at which the energy vanishes. If it is kind of
saddle points, then the high energy, fixed angle scattering amplitude could
fall off powerly. Indeed, we find that when $x_2$, $z_3$ and $z_4$ are off from
the boundary, the deviation of ${\cal E}$ is
\eqn\devia{\delta {\cal E}=s\delta x_2+(t+4)\delta r_3+ (u+4)\delta r_4,}
where we used the relation $s+t+u=-8$. All $\delta x_2$, $\delta r_3$ and
$\delta r_4$ are negative. To have a positive deviation, we would require
all Mandelstam variables be negative, so this is a nonphysical region.
It is meaningless to do integration over these deviations, given divergent
factors in \four. These factors come from the on-shell condition of
the tachyon, we can get rid of them by considering massless or massive
excitations. Let us now discuss four point amplitude of massless tensor
states. Let $\xi^{(i)}_{\mu_i\nu_i}$ be the polarization tensor of the i-th
massless tensor state, the scattering amplitude is
\eqn\tensor{A_4={1\over {\cal N}}\int \prod \xi^{(i)}_{\mu_i\nu_i}d^2z_i
\langle\partial X^{\mu_1}\overline{\partial}X^{\nu_1}\dots \partial X^{\mu_4}
\overline{\partial}X^{\nu_4}e^{i\sum p_iX(z_i)}\rangle.}
There are many possible contractions among operators in this case. One of these
is
$$\langle\partial X^{\mu_1}\partial X^{\mu_2}\rangle\langle\partial X^{\mu_3}
\partial X^{\mu_4}\rangle\langle\overline{\partial}X^{\nu_1}\overline{\partial}
X^{\nu_2}\rangle\langle\overline{\partial}X^{\nu_3}\overline{\partial}
X^{\nu_4}\rangle e^{-{\cal E}},$$
where ${\cal E}$ is given by \bizz\ provided we replace $s+4$ by $s$ etc.
The above contraction gives a factor
$${1\over |z_1-z_2|^4|z_3-z_4|^4}.$$
It is divergent when $z_1$ approaches $z_2$ or $z_3$ approaches $z_4$. Fix
conformal invariance in \tensor\ by letting $z_1=0$ and $z_2=x_2$. Again
the divergence from $x_2\rightarrow 0$ is due to the presence of tachyon.
Other divergence disappears if we properly extend Mandelstam variables to
a nonphysical region. However, when all $x_2$, $z_3$ and $z_4$ are
on the boundary, the divergence from $z_3\rightarrow z_4$ on the boundary can
not be removed in this way, since ${\cal E}$ becomes momentum-independent.
This divergence can be removed by adding a counter term on the boundary.
Thus we find that to have this theory well-defined (ignoring tachyon), we need
to modify Dirichlet conditions by certain regularization on the boundary.
We note that similar divergence also appears in scattering amplitudes
of closed string states if there are Neumann boundaries.
There are also contractions convergent when two vertex operators approach
each other. In these contractions, there are factors $(1-z_i\bar{z}_j)^{-2}$.
To see that there is divergence from this factor when both $z_i$ and $z_j$
are close to the boundary, we map the unit disk to the upper half plane.
Now near the boundary
$$\int {d^2z_id^2z_j\over (z_i-\bar{z}_j)^2}=\int{dy_idy_jdx_idx_j
\over(i+(x_i-x_j))^2}.$$
In the last equality we rescale $x_i$ by a factor $y_i+y_i$. So there is no
divergence when $x_i\rightarrow x_j$.

We use ${\cal E}$ to calculate the
high energy, fixed angle behavior. Again the deviation of ${\cal E}$ from its
value on the boundary is
$$\delta {\cal E}=s\delta x_2+t\delta r_3 +u\delta r_4.$$
Assuming that the real parts of $s, t, u$ are negative, integral over
deviations
gives us $1/(stu)$. Note that, there is no extra power in $s$ from the factor
$(1-z_i\bar{z}_j)^{-2}$, as should be clear from the above discussion of this
factor. This result was obtained previously by Green \mbg. Other
contractions occur in \tensor\ give the same high energy contribution.
It is easy to see that when all the Mandelstam variables are large, the energy
becomes large off the boundary, so $\hbox{exp}(-{\cal E})$ oscillates fast
for a properly chosen region of the Mandelstam variables. We
summarize that in the Dirichlet string theory, the high energy, fixed angle
scattering is governed by a ``saddle circle'', the Dirichlet boundary.
This is why its fall-off is power-like. Physically, one can imagine that
when high energy scattering states approach together, they suddenly undergo
point-like interaction, as enforced by the Dirichlet boundary.

\subsec{Annulus amplitudes and general amplitudes}

Annulus amplitudes can be calculated using the Green function given in \annu.
There is only one conformal Killing vector in this case, generating
rotations on the annulus. The volume of this simple conformal group is finite,
there is no necessity to fix it. The amplitude for tachyons is
\eqn\form{A_n=\int\prod_i d^2z_i {da\over a}\eta^{-24}(a^2)e^{-2\pi\sum_{ij}
p_ip_jG(z_i,z_j)}.}
Again, due to the nature of tachyon, when one of $z_i$ approaches the outer
boundary, there is divergence. Similarly, when one of $z_i$ approaches to
the inner boundary, there is also divergence, because of the $n=1$ term
in the last
term of \annu. We can consider massless tensor states to get rid of these
divergences. Once again, when all $z_i$ are on the outer boundary, the
energy in the exponential in \form\ vanishes. The fluctuation of the energy
is no longer linear in fluctuations of punctures. It is quadratic in $\delta
r_i$:
$$\eqalign{\delta {\cal E}&=\sum_{i\ne j}p_ip_jg(\theta_i,\theta_j)\delta
r_i\delta r_j,\cr
g(\theta_i,\theta_j)&={1\over 2\hbox{sin}^2(\theta_i-\theta_j)/2}+2\sum_{n=1}
\left({a^{2n}e^{i(\theta_i-\theta_j)}\over (1-a^{2n}e^{i(\theta_i-\theta_j)}
)^2}+c.c\right),}$$
when $p_i^2=0$. Integration over $\delta r_i$ results in a determinant
\eqn\deter{\left(\hbox{det}(p_ip_jg(\theta_i,\theta_j))\right)^{-1/2}.}
This expression is not singular when two $\theta$'s come together.
We would encounter a similar divergence arising from certain contraction
of oscillators
in the vertex operators of massless tensor states, when two vertex operators
approach to each other on the boundary. This divergence is independent of
moduli, hence can be regularized by adding a boundary counter term as in the
disk case. We specify to the case $n=4$. After performing integration
over $\theta_i$, we expect that the
amplitude obtained from \deter\ is proportional to $s^{-2}$ at fixed angle.
It is surprising that this decreases more slowly than the amplitude on the
disk, which is proportional to $s^{-3}$.

Moduli integration in \form\ is divergent at both ends $a=0$ and $a=1$.
The divergence at $a=0$ has the same origin as the divergence in the
annulus partition function. To remove this divergence, one may need to
remove the tachyon as well as the massless dilaton from the spectrum.
There is another divergence at $a=1$. To see this, recall that when
$a\rightarrow 1$, $\eta^{-24}(a^2) \rightarrow (\hbox{ln}a/\pi)^{-12}
\hbox{exp}(-{2\pi^2\over\hbox{ln}a})$ which blows up at $a=1$. This
formula follows from the modular transformation of the eta function.
In order to extract the asymptotic behavior of the Green function, we use
$$G(z,w)={-1\over 4\pi}\hbox{ln}|{\theta_1(\nu_1|\tau)\over \theta_1
(\nu_2|\tau)}|^2,$$
where $\hbox{exp}(2\pi i\nu_1)=z/w$, $\hbox{exp}(2\pi i\nu_2)=z\bar{w}$
and $\hbox{exp}(\pi i\tau)=a$.
Using the modular transformation for the theta function, we find
$$G(z,w)\rightarrow -{1\over 4\pi}\left(\hbox{ln}|e^{{2\pi^2(\nu_1^2-\nu_2^2)
\over
lna}}|+\hbox{ln}|{\theta_1(-{i\pi\nu_1\over lna}|-{i\pi\over
lna})\over \theta_1(-{i\pi\nu_2\over lna}|-{i\pi\over ln
a})}|^2\right).$$
When $a\rightarrow 1$, the above formula yields
$$G(z,w)\rightarrow {1\over 4\pi \hbox{ln}a}\left((\hbox{ln}|z/w|)^2-(
\hbox{ln}|zw|)^2\right).$$
As a double check, the above formula is symmetric in $z$ and $w$ and
vanishes when $|z|=1$. This formula tell us that $G(z,w)$ is always
negative when $a\rightarrow 1$. So it can be used to regularize the
integral over $a$ in \form. In doing this, one only let positions of
vertex operators keep away from the boundary slightly. In the case of
four point amplitude with high energy and fixed angle, let $p_ip_j
=sg_{ij}$, $g_{ij}$ depends on the angle. Now the cut-off on $r_i=
|z_i|$ is imposed in such a way that
\eqn\cutoff{-{1\over\hbox{ln}a}\left(2\pi^2+{s\over 2}\sum_{ij}
g_{ij}((\hbox{ln}|z_i/z_j|)^2-(\hbox{ln}|z_iz_j|)^2)\right)<0.}
So the cut-off is of order $\delta r_i\sim 1/\sqrt{s}$. Note that,
when $s>0$, $g_{ij}$ must be positive definite in order to satisfy
\cutoff. In adopting this cut-off, we do not derive the logarithmic
corrections as found in \ref\yang{Z. Yang. UR-1288/ER-40685-737.}

What happens when all $z_i$ are on the inner boundary? The Green
function $G(z_i,z_j)$ is a constant $-1/(4\pi) \hbox{ln}a^2$. For this reason,
the energy becomes momentum-independent too. Again the fluctuation of the
energy is quadratic in $\delta r_i$, and integration over $\delta r_i$ yields
a power $s^{-2}$. From studying
the Green function, we find that when one of the argument on the outer
boundary, and the other on the inner one, it is vanishing too. The leading
terms in the fluctuation of the energy is linear in deviations of those
operators on the outer boundary. So when there is only one operator close
to the outer boundary and the rest are on the inner one, contribution
from this configuration to the amplitude is proportional to $s^{-5/2}$,
decreases faster than the contribution when all operators are on the
same component. When two operators are on the outer boundary, and the
other two are on the inner one, contribution from this configuration is
of order $s^{-3}$. When three operators are on the outer boundary,
contribution is also of order $s^{-3}$. In conclusion, we find that the
leading power behavior
of the annulus amplitude is $s^{-2}$, and the term comes from configurations
when all vertex operators are on the same boundary.
In all, when all the vertex operators are on the boundary, even they are on
different components, power-like contribution arises from integration over
fluctuations.

The argument for the power-like fall-off of four point amplitudes on the disk
and the annulus generalizes to n-point amplitudes on an arbitrary surface
with at least one component of
the boundary. It is easy to see that the Green function $G(z,w)$ on such
a surface is a constant when both $z$ and $w$ are on the boundary. We already
showed that when one of the variables is on the boundary, $G(z,\theta_i)$ is
independent of $\theta_i$ (it may still depend on $z$).
Let $z$ approach
a component of the boundary, by continuity, $G(\theta_j,\theta_i)$ is
independent of $\theta_j$ too.
Now if a subset of vertex operators is on one component, and the rest
is on another component, the energy becomes independent of the Mandelstam
variables. Integration over fluctuations from these two components yields
power-like amplitudes. These configurations are dominant over other region.
When the Mandelstam variables are all large, the absolute value
of the energy becomes large in off-boundary, and $\hbox{exp}(-{\cal E})$
oscillates fast.

\newsec{Finite temperature partition function}

Partition function at high temperature is always a good probe of the nature
of a quantum field theory as well as a string theory. It is often
easier to calculate than scattering amplitudes. In the string theory,
mass degeneracy grows exponentially, so there exists
a limiting temperature called the Hagedorn temperature \ref\hag{K. Huang
and S. Weinberg, Phys. Rev. Lett. 25 (1970) 895; S. Fubini and G. Veneziano,
Nuovo Cim. 64A (1969) 1640; R. Hagedorn, Nuovo Cim. Suppl. 3 (1965) 147.}.
Above this temperature, temperature loses its meaning, one has to replace
the  canonical ensemble by the micro-canonical ensemble \ref\micro{S.
Frautschi, Phys. Rev. D3 (1971) 2821; R.D. Carlitz, Phys. Rev. D5 (1972)
3231.} \ref\chung{N. Deo, S. Jain and C.-I Tan, Phys. Lett. B220 (1989) 125;
Phys. Rev. D40 (1989) 2626.}.
The partition function, to the leading order (on the torus) in string theory,
was calculated in a path integral approach in \ref\jpl{J. Polchinski,
Commun. Math. Phys. 104 (1986) 37.}, and was subsequently shown to
be modular invariant in \ref\minv{B. MacClain and B. Roth, Commun. Math.
Phys. 111 (1987) 539; K.H. O'Brien and C.-I. Tan, Phys. Rev. D36 (1987)
1184.}. The high temperature behavior, in case if one can analytically
extend the low temperature result to high one, was discussed in \ref\awit
{J. Atick and E. Witten, Nucl. Phys. B310 (1988) 291.} and found to be
universal to all orders. The major consequence of this investigation is
that the free energy per unit volume is proportional to $T^2$.
This implies that in a string theory, there are far fewer degrees of freedom
than in a quantum field theory, as one expects in the latter the partition
function would go as $T^4$ at high temperature. Now in QCD, as energy
or energy density goes high, the point-like structure dominates and
the partition function should behave dramatically differently from the usual
string theory. As a calculation of Polchinski shows, a winding mode develops
into tachyon at high temperature and its mass square goes as $-g^2N/\beta^2$,
where $g^2$ is the QCD coupling constant, $N$ is the rank of $SU(N)$ gauge
group and $\beta$ the inverse of temperature. A natural question to
ask is what will happen to the finite temperature partition function and
spectrum in the
Dirichlet string theory. Green in \mbgg\ calculated two point function of
the winding tachyon state on the disk at finite temperature, and found a
similar behavior as predicted by Polchinski.

In this section, we will discuss the finite temperature partition function
at an arbitrary
genus with an arbitrary number of boundaries, based on the approach presented
in section 2. The high temperature behavior can be easily read off from the
formal approach. Moreover, we shall give an explicit expression for the
partition function on the annulus. The major difference we find in the
Dirichlet theory from in the usual theory is that already starting from
genus zero, the partition function depends on temperature, and infinite
number of boundaries dominates if one takes the boundary coupling constant to
be that in QCD. This is true also at higher orders. It may appear surprising
that there is temperature dependence of the free energy at the tree level.
All components of the boundary on the sphere are mapped to spacetime points,
so there could be no nonvanishing winding modes. Indeed there is no
contradiction. As we shall see, the formal expression of the free energy
has a divergence. This divergence suppresses contribution from nonvanishing
winding modes. However, the divergence is independent of moduli, and can be
regularized by adding a boundary counter term. This mild modification
can be thought of as opening little windows on the world sheet, as one would
expect from studying QCD. So we find the following intriguing possibility.
The role of Dirichlet boundaries is twofold. On the one hand, it enforces
high energy particles suddenly undergo point-like interaction when they
approach together. On the other hand, it helps open little windows on the
world sheet so that the tree level free energy becomes $T$-dependent at
high enough $T$.

We start with arguing that our approach in section 2 directly applies
here. Note that at finite temperature, spacetime is Euclidean with a
compactified time. The circumference of the compactified time is $\beta$.
We should be careful in inserting the delta function into the path integral
to enforce Dirichlet boundary condition. Instead of inserting a factor
$\delta(X^0(\theta)-x^0)$ on the boundary, we have to insert $\sum_{n}
\delta(X^0(\theta)-x^0-n\beta)$, in order to take care of periodicity in
time. Again we introduce a boundary field $\rho^0(\theta)$. The delta function
is written as
$$\sum_n\delta(X^0(\theta)-x^0-n\beta)=\sum_n\int[d\rho]e^{-{i\over
4\pi}\int\rho(\theta)(X^0(\theta)-x^0-n\beta)ds}.$$
Upon inserting this delta function into the path integral, we perform the
integration over $x^0$. Now the range of $x^0$ is $[0, \beta]$. The sum in
$n$ in the above formula can be absorbed into the integration, and the range
of $x^0$ is changed to $(-\infty, \infty)$, we then get the same expression
as in \def\ (there is no insertion of vertex operators, if we are concerned
with the partition function only). We do the same thing for spatial
dimensions as we did in section 2. Thus, we end up with the same prescription
as in sect. 2.

To proceed further, we need to find the classical solution satisfying
\eqn\repeat{-\Delta X_{cl}=0, \quad \partial_n X_{cl}(\theta_i)=-i\rho
(\theta_i),}
where we introduced index $i$ to label components of the boundary. If we
deal with the zero temperature partition function, we use the solution
obtained in sect. 2. However, we are dealing with a compactified time,
periodicity in the time direction must be taken care of. Suppose the genus
of the surface is $g$, and there are $L$ components of the boundary. As
usual we label those homological inequivalent circles on the surface by
$a_i$ and $b_i$ \ref\rie{see any book on Riemann surfaces.}. There are more
such circles we take as $L-1$ of $L$ components of the boundary. Apparently,
if we use the formula $X(z)=-i\sum_i \int\chi(z,\theta_i)\rho(\theta_i)ds_i$,
we would end up with a solution without winding numbers along those nontrivial
homological circles. Note that, since we have no other boundary conditions
on $X_{cl}$ except that in \repeat, there could be nonzero winding numbers
along all nontrivial homological circles. Let us denote these winding numbers
as $m_i$ along $a_i$ and $n_i$ along $b_i$, $i=1,\dots, g$, and $l_i$
along the $i$-th component of the boundary, $i=1,\dots, L-1$. The classical
solution then consists of two pieces, one is the one satisfying \repeat\
without winding numbers, another is the solution to
\eqn\usual{-\Delta X(z)=0,\quad \partial_n X(\theta_i)=0}
with winding numbers $(m_i,n_i, l_i)$ in $X^0$ direction. Denote this piece
by $X^{(2)}_{cl}$.
This is what we would obtain in the Neumann open string theory. To obtain
such a solution, we use a trick. Consider a mirror image  of the Riemann
surface under consideration, the images of $a_i$, $b_i$ and $l_i$ are denoted
by $\bar{a}_i$, $\bar{b}_i$ and $\bar{l}_i$. Now we glue the Riemann surface
to its mirror image along components of boundaries, such that $l_i$ is glued to
$\bar{l}_i$. We obtained a new Riemann surface of genus $G=2g+L-1$ without
boundary. The solution $X_{cl}^{(2)}$ is just the classical solution on this
new Riemann surface in which the winding number along a holomogical circle
is the same as along its mirror image. There are a number $G=2g+L-1$ of
independent winding numbers. Note that there are additional nontrivial
homological circles on the new Riemann surface. Winding numbers are necessarily
zero along these circles, because any of these circles consists of half
on the ordinary surface and half on its image, so map from the first half to
$X^0$ is traced back on the second half. We denote by $A_i$ and $B_i$
the homological circles on the new Riemann surface. In particular,
$A_i=a_i$, $B_i=b_i$, $A_{2g+1-i}=\bar{a}_i$ and $B_{2g+1-i}=\bar{b_i}$
when $i=1,\dots, g$, and $A_{2g+i}=l_i$ when $i=1,\dots, L-1$. Pick a basis of
holomorphic differentials $\omega_i$ such that
$$\int_{A_i}\omega_j=\delta_{ij},\quad \int_{B_i}\omega_j=\Omega_{ij},$$
where indices run from $1$ to $2g+L-1$. The period matrix $\Omega$, though
defined for the new surface, depends only on moduli of the ordinary surface.
The classical solution $X_{cl}^{(2)}$ to \usual\ is written in terms of
the period matrix as
\eqn\sec{X_{cl}^{(2)}={i\over 2}\beta\int^z \sum\left((\hbox{Im}\Omega)^{-1}
_{ij}(\overline{\Omega}_{jl}M_l-N_j)\right)\omega_i+c.c.}
where $c.c.$ abbreviates complex conjugate. The winding numbers in \sec\ are
$$\eqalign{M_i&=M_{2g+1-i}=m_i, \quad N_i=N_{2g+1-i}=n_i \quad
\hbox{when} \quad i=1,\dots, g\cr
M_{2g+i}&=l_i, \quad N_{2g+i}=0, \quad \hbox{when} \quad i=1,\dots L-1.}$$
Note that \sec\ gives a nonzero piece only for $X^0(z)$. Now substituting
$X_{cl}=
-\int \chi(z,\theta_i)\rho(\theta_i)ds_i+X_{cl}^{(2)}$ into action
\act\ and taking care of multiple-valueness of $X^0(z)$, we find the classical
action
\eqn\cact{\eqalign{S_{cl}&={1\over 8\pi}\int \rho(\theta_i)\chi(\theta_i,
\theta_j) \rho(\theta_j)ds_ids_j+{i\over 4\pi}\int\rho(\theta_i)X_{cl}^{(2)}
(\theta_i) ds_i\cr
&+{\beta^2\over 16\pi}(\overline{\Omega}_{il}M_l-N_i)(\hbox{Im}\Omega)^{-1}
_{ik}(\overline{\Omega}_{kj}M_j-N_k),}}
where we have used the fact that $\partial X_{cl}^{(2)}$ is single-valued
on the surface. The last term is equal to the half of the classical action
of the solution $X_{cl}^{(2)}$ on the new surface. Next we integrate out
field $\rho$ and obtain
\eqn\cacti{S_{cl}={1\over 8\pi}\int X^{(2)}_{cl}(\theta_i)\chi^{-1}(\theta_i,
\theta_j)X^{(2)}_{cl}(\theta_j)ds_ids_j+{\beta^2\over 16\pi}(\overline{\Omega}
_{il}M_l-N_i)(\hbox{Im}\Omega)^{-1}_{ik}(\overline{\Omega}_{kj}M_j-N_k).}
We note that the first term on the r.h.s. of the above formula is proportional
to $\beta^2$, just as the second term. In addition, the first term is also
quadratic in winding numbers. There is potential divergence in this term
when $\theta_i$
approaches $\theta_j$ (on the same component of the boundary), as we shall
see in the annulus case. Now we denote winding numbers collectively
by a row matrix $Q=(m_i,n_i,l_i)$, then the above formula may be succinctly
written
$$S_{cl}(Q)={\beta^2\over 8\pi}QAQ^{t},$$
where $A$ is a $G\times G$ matrix independent of temperature. Notice that our
result concerning the classical action is similar to what obtained in the
usual string theory \minv. Finally, the partition function on the genus $g$
surface with $L-1$ components of boundary is
\eqn\parti{F/V=-\int [dm]\sum_Qe^{-S_{cl}(Q)},}
where $F/V$ is the free energy per unit volume at this order, $[dm]$ is the
measure on the moduli space, which can be calculated once the Green function
discussed in sect. 2 is known. Note that $Z=-\beta F$ is the string partition
function, we have dropped a factor $\beta$ in \parti\ since integration over
$X^0$ zero mode gives a factor $\beta$.

We come to the conclusion that even at the tree level when $g=0$, there could
be nontrivial $T$-dependence of the free energy if $L>1$. So starting
from the annulus, the free energy becomes nontrivial. This is in contrast to
the closed string theory where there is no nontrivial free energy at the tree
level, at least below the Hagedorn temperature \awit. In our case, although
boundaries are mapped to single spacetime points, our approach of introducing
the $\rho$ field allows nonzero winding numbers on each component of the
boundary, and these winding modes contribute. This is a rather surprising
result. In the open string theory, winding modes are also allowed along
boundaries. The difference is that, adding a boundary in the open string
theory increase a open string loop, so at the tree level there is no
dependence of the free energy on $T$. We shall demonstrate our  calculation
on the annulus in the end of this section. We now turn to discussing the
high temperature behavior of \parti.

We believe that there also exists the Hagedorn temperature in the present
theory, and the notion of temperature may lose its meaning in some
region above it. On the other hand, if the Dirichlet string has anything to
do with the QCD
string, one expects that at high enough temperature, the definition of
temperature exists as one essentially is dealing with the quark-gluon plasma.
We assume that \parti\ can be analytically extended to high
$T$ for a given topology. We follow \awit\ to calculate the high $T$
behavior of \parti. We approximate the discrete sum by integration over
continuous variables $Y=\beta Q$. If $A$ is nonsingular, the sum in \parti\
is approximately
$$\left({\sqrt{8}\pi\over\beta}\right)^{2g+l-1}(\hbox{det}A)^{-1/2}.$$
This expression grows fast at high $T$ when either $g$ or $l$ or both
is large. Remember
that we have to multiply the free energy by a factor $\lambda_c^{2g-2}$
from closed string loop interaction, where $\lambda_c$ is the closed string
coupling constant. Also, each component of the boundary is accompanied with
a factor
$\lambda_o\sim g^2$, where $g$ is the QCD coupling constant. Green argued
that unitarity does not impose any relation between $\lambda_c$ and
$\lambda_o$. So the right formula for the integrand in the free energy is
\eqn\righ{\left({\sqrt{8}\pi\lambda_c\over\beta}\right)^{2g-2}\lambda_o^l
\left({\sqrt{8}\pi\over\beta}\right)^{l+1}(\hbox{det}A)^{-1/2}.}
Following \awit, we assume $\lambda_c^2T^2$ is fixed at high $T$. The above
expression
still grows fast when $l$ is large, since the coupling constant in QCD
decreases as $1/\hbox{ln}T$ at high $T$. We conclude that the free energy
at each loop is dominated by large $l$. To get a sensible result, one then has
to sum over all $l$. This is a hard problem, since it is not easy to perform
moduli integration in \parti. We point out that it is questionable to
assume that $\lambda_c T$
is fixed at high temperature, if one presumes that topological expansion in
the Dirichlet string theory is similar to $1/N$ expansion in QCD. In QCD,
the rank $N$ of the gauge group can not get renormalized.

In the rest of this section, we calculate \parti\ on the annulus. In this
simplest case, we do not need to use the trick of gluing the mirror of
the annulus to itself. The classical solution with winding number $n$ along
one of the boundary is ${n\beta}/(2\pi)\theta$. The action of this solution
is just
\eqn\second{{n^2\beta^2\over 16\pi^2}\hbox{ln}a^{-1},}
where $a$ is the inner radius of the annulus. It is straightforward to
calculate the first term in the classical action
in \cacti, induced by integration over the $\rho$ field. Using our result
about $\chi^{-1}$ in section 2, we find the first term in \cacti
\eqn\recov{{n^2\beta^2\over 4\pi}(\chi_{11}(2\pi, 2\pi)-\chi_{12}(2\pi,2\pi)),}
where $\chi_{11}$ and $\chi_{12}$ as defined in section 2. There is a
divergence in $\chi_{11}(2\pi, 2\pi)$, reflecting the fact that when
$\theta$ and $\theta'$ in $\chi_{11}(\theta,\theta')$ approach together,
there is a singularity which has to be regularized. After dropping out this
$a$-independent divergent term, the finite part of \recov\ reads
\eqn\first{{n^2\beta^2\over 4\pi^2}\hbox{ln}\left(\prod_{m=1}^\infty ({1-
a^{2m-1}\over
1-a^{2m}})^2\right)={n^2\beta^2\over 4\pi^2}\hbox{ln}\left({a^{1/4}\theta_3
(1/2|-i\hbox{ln}a/\pi)\over \eta^3(a^2)}\right),}
where $\theta_3(\nu|\tau)$ is one of the Jacobi theta functions, and $\eta(q)$
is just $q^{1/24}\prod (1-q^n)$. It is interesting to note that the $a^{1/4}$
factor in the logarithm in the above formula cancels out when we add \second\
to \first. Taking into account the measure on the moduli space, the free
energy on the annulus reads
\eqn\free{F/V=-\int_0^1{da\over a}\eta^{-24}(a^2)G(\beta,a),}
where
$$G(\beta, a)=\sum_{n=0}^\infty \hbox{exp}\left(-{n^2\beta^2\over 4 \pi^2}
\hbox{ln}{\theta_3(1/2|-i\hbox{ln}a/\pi)\over \eta^3(a^2)}\right)$$
When $\beta\rightarrow \infty$, we recover the zero temperature partition
function. We learn from this example that the free energy has a temperature
dependence even at the tree level.

We have omitted the divergence part, as we assumed it be removed by
adding a boundary counter term. If we put a cut-off on $\theta-\theta'$
in function $\chi_{11}(\theta,\theta')$, say $\epsilon$, then the
divergent part in \recov\ looks like
$${n^2\beta^2\over 4\pi^2}\hbox{ln}\epsilon^{-1}.$$
It is independent of the moduli parameter $a$. This term makes $S_{cl}$
blow up if $n\ne 0$, and suppresses contributions of nonvanishing winding
modes. Subtracting this divergence then amounts to opening windows on
the world sheet, so that the free energy receives contributions from
nonzero winding modes along boundaries. We suspect that this subtraction
of divergence is the same thing we had to do to make scattering amplitudes
well defined, as we discussed in section 3.

It is good to have an exact formula such as \free. However, we can not
read off much from \free. First, as we noticed
previously, it is necessary to sum over all numbers of boundaries, even
at the tree level, to have a reliable formula of the free energy. Second,
to learn what the spectrum at a finite temperature looks like from the free
energy, we have to calculate at least the torus with one boundary.

\newsec{Discussion}

We have shown in this paper that Dirichlet strings display some desirable
properties of QCD at high energy and high temperature, using a simple
approach to the model. Perhaps the most attractive feature is the possibility
to give $T$-dependence to the sphere free energy. One expects from QCD that
the dependence appears only above a certain temperature, where the picture
of a perfect surface without holes breaks down. We have not addressed the
problem of determining this temperature in this paper. It is worth to note
that, since power-like behavior arises at any order with any number of
boundaries, it is necessary to sum up all surfaces to have a reliable answer.

{}From a purely theoretic point of view, there are several serious problems
with this string theory. First, as QCD makes perfect sense in 4 dimensions,
there must be an effective 4D string theory. The Dirichlet string lives in
26 dimensions, at least perturbatively. A way to reduce spacetime dimensions
is to introduce a fermionic Dirichlet string theory, in which there is
no spacetime supersymmetry. Another outstanding problem is the existence of
a tachyon. Maybe one should not worry about this too much, because mass
spectrum is determined by interaction. To determine it, one perhaps has to
calculate zero temperature partition function on the torus with all possible
numbers of Dirichlet boundaries and sum it up. Our work presented here may
serve as a
starting point to address this problem. We would like to point out another
interesting model proposed in \zg\ which shares nice properties of the simplest
Dirichlet string theory studied here. In particular, mass spectrum is
modified by interaction in that model too.
{}From a more physical point of view, to have a good QCD string theory, we
have to introduce quarks living on
Neumann-type boundaries. This is also an open problem. We expect that high
energy scattering amplitudes of mesons also display power fall-off at high
energy.

Finally, we note that in the string theory of the 2D pure Yang-Mills theory
\qcd, maps with some handles and tubes mapped to spacetime points are included,
and contribute significantly. This is similar to the spirit of the Dirichlet
string theory in which boundaries are mapped to points. The relation
between the Dirichlet string theory and the 2D pure Yang-Mills string theory
deserves careful study.

\vskip0.7cm
\noindent {\bf Acknowledgments}

I wish to thank P. Mende, C.-I. Tan and W. Zhao for helpful discussions.
This work was supported by DOE contract DE-FG02-91ER40688-Task A.

\listrefs
\end